\documentclass[runningheads]{llncs}
\usepackage{bbding}
\usepackage{tikz}
\usepackage{calc}
\usepackage{graphicx}
\usepackage{amsmath}
\usepackage{amsfonts}
\begin{document}
\title{Learning to Address Intra-segment Misclassification in Retinal Imaging}
%
%
\author{Yukun Zhou\inst{1,2,3}, Moucheng Xu\inst{1,2}, Yipeng Hu\inst{1,2,5}, Hongxiang Lin\inst{1,3,6}, Joseph Jacob\inst{1,7}, Pearse A. Keane\inst{3,8}, and Daniel C. Alexander\inst{1,4}
}

%
\authorrunning{Y. Zhou et al.}
%
\institute{Centre for Medical Image Computing, University College London, London, UK \email{yukun.zhou.19@ucl.ac.uk}\\
\and
Department of Medical Physics and Biomedical Engineering, UCL, London, UK
\and
NIHR Biomedical Research Centre at Moorfields Eye Hospital, London, UK \and 
Department of Computer Science, University College London, London, UK
\and
Wellcome/EPSRC Centre for Interventional and Surgical Sciences, London, UK \and
Research Center for Healthcare Data Science, Zhejiang Lab, Hangzhou, China \and
UCL Respiratory, University College London, London, UK
\and
UCL Institute of Ophthalmology, University College London, London, UK
}
\maketitle              
\begin{abstract}

Accurate multi-class segmentation is a long-standing challenge in medical imaging, especially in scenarios where classes share strong similarity. Segmenting retinal blood vessels in retinal photographs is one such scenario, in which arteries and veins need to be identified and differentiated from each other and from the background. Intra-segment misclassification, i.e. veins classified as arteries or vice versa, frequently occurs when arteries and veins intersect, whereas in binary retinal vessel segmentation, error rates are much lower. We thus propose a new approach that decomposes multi-class segmentation into multiple binary, followed by a binary-to-multi-class fusion network. The network merges representations of artery, vein, and multi-class feature maps, each of which are supervised by expert vessel annotation in adversarial training. A skip-connection based merging process explicitly maintains class-specific gradients to avoid gradient vanishing in deep layers, to favor the discriminative features. The results show that, our model respectively improves F1-score by 4.4\%, 5.1\%, and 4.2\% compared with three state-of-the-art deep learning based methods on DRIVE-AV, LES-AV, and HRF-AV data sets. Code: \textbf{https://github.com/rmaphoh/Learning-AVSegmentation}

\end{abstract}
\section{Introduction}

\begin{figure}[t]
\centering
\includegraphics[width=0.8\columnwidth]{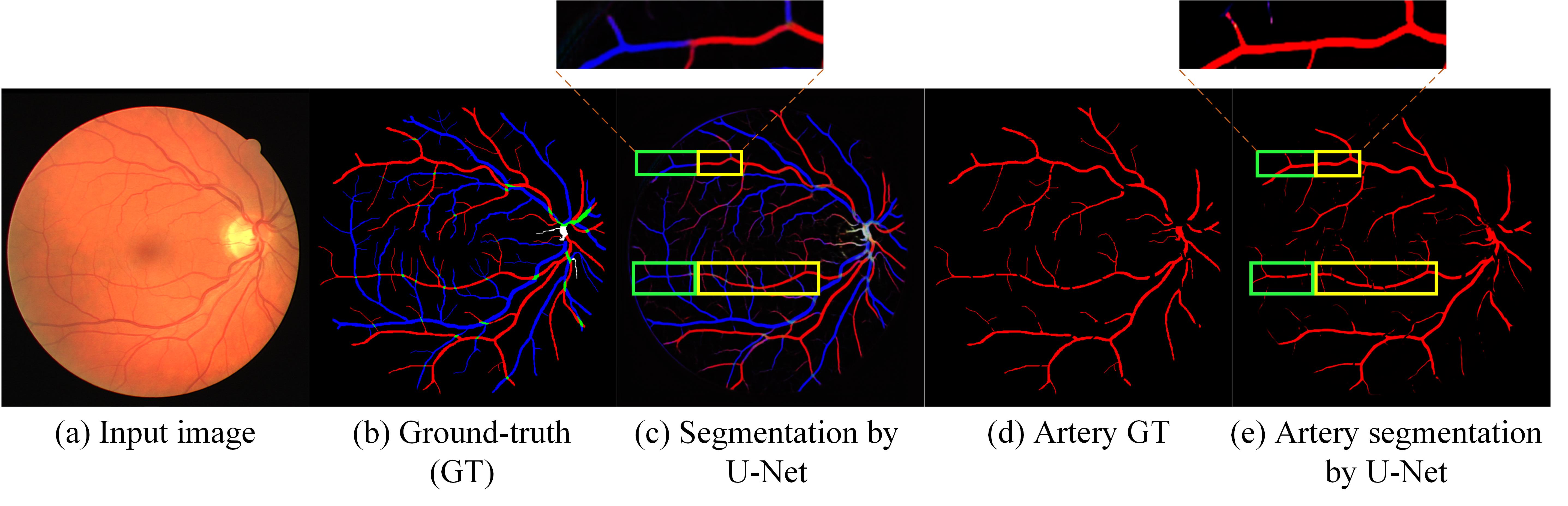} 

\caption{Illustration of retinal vessel segmentation maps highlighting common errors arising at vessel intersections. In (c), the green boxes highlight two partial arteries that are misclassified as veins in a multi-class segmentation algorithm, while correctly classified as arteries in the binary segmentation task (e).} 

\label{fig1}
\end{figure}

\begin{figure}[t]
\centering
\includegraphics[width=0.7\columnwidth]{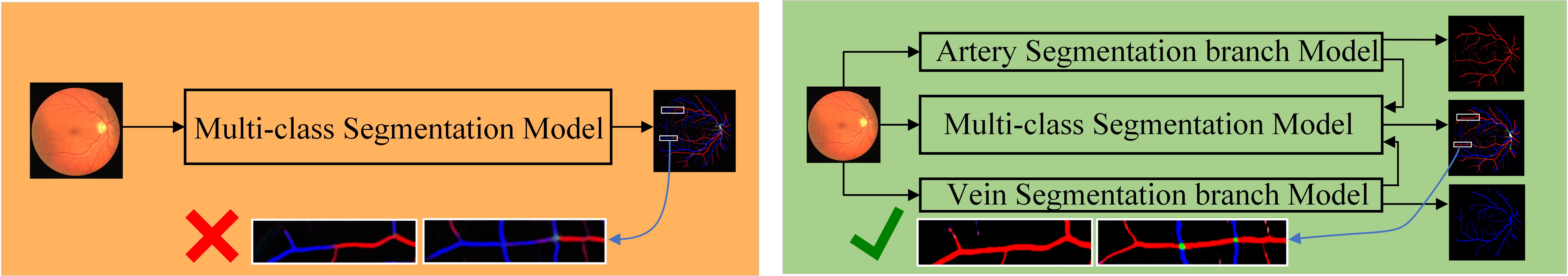} 
\caption{Strategy of the binary-to-multi-class fusion.} 
\label{fig2}
\end{figure}

Measurements of retinal vessel morphology are vital for computer-aided diagnosis of ophthalmological and cardiovascular diseases, such as diabetes, hypertension, and arteriosclerosis \cite{fraz2012blood,wong2001retinal}. Retinal vessel segmentation on fundus images plays a key role in retinal vessel characterisation, by providing rich morphological features that are sensitive for diagnosis. However, manual annotation of retinal vessels is highly time-consuming, thus automated vessel segmentation is in high demand to improve the efficiency of the diagnosis, with or without computer assistance, of a wide range of diseases. 

In automatic multi-class retinal vessel segmentation, i.e. segmenting veins and arteries from retinal fundus images, there are two main categories, namely, graph-based methods and feature-based methods. Graph-based methods utilise the vessel topological and structural knowledge to search for the vessel tree structures and classify them into artery and vein \cite{dashtbozorg2013automatic,wang2020multi,estrada2015retinal,srinidhi2019automated,xie2020classification,zhao2019retinal}. Feature-based methods design various feature extraction tools to discriminate the vessel category~\cite{huang2018artery,mirsharif2013automated}. The methods that have achieved state-of-the-art performance have come from end-to-end deep-learning-based models ~\cite{hemelings2019artery,xu2018simultaneous,wu2020nfn+,fu2016deepvessel,ronneberger2015u,galdran2019uncertainty,meyer2018deep,wang2019retinal}. However, the intra-segment misclassification that occurs around inter-class vessel intersections, as illustrated in Figure \ref{fig1}, is a well-recognised common issue across all the deep-learning-based models.

In this paper, we focus on intra-segment misclassification in multi-class retinal vessel segmentation \cite{li2020joint}. More specifically, we observed and identified that a type of intra-segment misclassification frequently happens when the inter-class vessels intersect, as illustrated in Figure \ref{fig1}(c), and has been reported in~\cite{galdran2019uncertainty,ma2019multi,hemelings2019artery,xu2018simultaneous,meyer2018deep,wu2020nfn+,wang2019retinal,mou2019cs}. However, the binary segmentation, e.g., extracting only arteries from images, illustrated in Figure \ref{fig1}(e), was able to correctly identify the single class arteries, probably due to the simplified task that has less uncertainty pixel labels, such as those at inter-class vessel intersections. This has motivated a novel multi-class segmentation algorithm that decomposes the problem into a combination of multiple binary segmentation tasks followed by a binary-to-multi-class fusion step, as shown in Figure \ref{fig2}. This avoids direct learning on ambiguous inter-class vessel intersections. We thus hypothesise that the network specifically designed will substantially improve performance in local areas where inter-class vessel intersections confuse multi-class classification.

\begin{figure}[t]
\centering
\includegraphics[width=0.93\columnwidth]{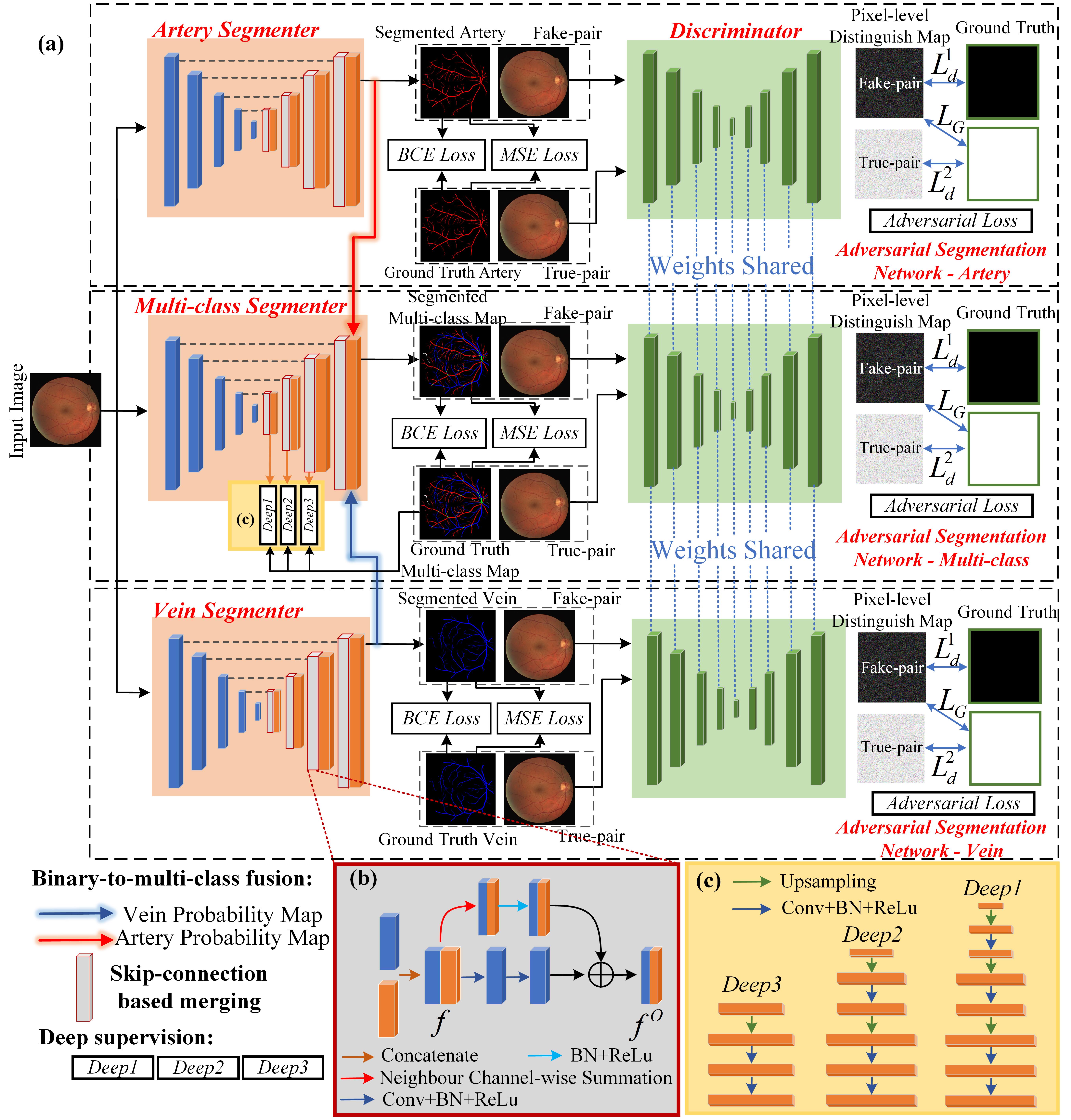} 
\caption{Flowchart of the binary-to-multi-class fusion network. The skip-connection based merging is highlighted in (b). The deep supervision structure is shown in (c). $L_{d}^{1}$,  $L_{d}^{2}$, and $L_{G}$ are the three loss objectives in adversarial loss.}
\label{fig3}
\end{figure}

We summarise our contributions as follows. First, we identified that the intra-segment misclassification happens at inter-class intersections in multi-class segmentation, while not in the binary. Second, we propose a deep learning model to address intra-segment misclassification, following the binary-to-multi fusion strategy. The flowchart is shown in Figure \ref{fig3}. Lastly, we report experiments including ablation studies on three real-world clinical data sets, to demonstrate the effectiveness of the proposed model.

\section{Methods}

We devise a deep learning based model, consisting of two components, an Adversarial Segmentation Network, as the backbone of the fusion network, and a Binary-to-multi-class Fusion Network. The Adversarial Segmentation Network detects and corrects higher-order inconsistencies between expert annotation maps and the ones produced by the generator, termed a segmenter here \cite{luc2016semantic,son2019towards}. The binary-to-multi-class fusion network merges the representations learnt from individual vessel types into a multi-class feature, to supplement the discriminative information of each single class surrounding the inter-class intersections.

\subsection{Adversarial Segmentation Network}

The Adversarial Segmentation Network compromises a segmenter and a pixel-level discriminator. The segmenter is a variant of U-Net while the discriminator is vanilla U-Net \cite{ronneberger2015u}. The input of segmenter is retinal fundus image and output is multi-class segmentation map. For discriminator, we concatenate the segmentation map and fundus image as input and obtain a pixel-level distinguish map as output, as shown in Figure 3.

\subsubsection{Skip-connection based merging process in segmenter.} This structure is employed to alleviate the gradient vanishing, as well as maintaining high resolution information. The structure is shown in Figure 3(b). We start with a concatenation of the up- and down-sampling feature maps at the same level into a combined feature $f\in\mathbb{R}^{W\times H\times N}$, where $W$, $H$, and $N$ respectively represent width, height, and channel of the spatial feature, which is divisible by output channel $C$. The merging process consist of a main path and a skip path. An operation in the main path $\phi:\mathbb{R}^{W\times H\times N}\rightarrow \mathbb{R}^{W\times H\times C}$ denotes two blocks, one of which comprises a convolution operation, batch normalisation, and activation function like ReLU in this case. Another operation in the skip path $\psi$ consists of batch normalisation and ReLU. Then the $k$th channel of the output feature $f^{O}$, denoted as $f_k^{O}$, is

\begin{equation}\label{eq1}
    f_k^{O} = \phi(f)_k+\psi\left(\Sigma_{i=1}^{N/C} f_{(k-1)N/C+i}\right),
\end{equation}

\noindent where $\sum_{i=1}^{N/C} f_{(k-1)N/C+i}$ indicates the Neighbour channel-wise summation, separately operating on the up- and down-sampling features.

\subsubsection{Adversarial training.} The adversarial training is employed to prompt the segmentation capability \cite{son2017retinal,chen2020tr}. Specifically, there are two reasons to adapt a U-Net as discriminator. First, the adversarial loss is pixel-wise, thus every pixel value in a segmented map mimics that in the experts annotation, thereby obtaining clear and sharp segmentation with high resolution information \cite{son2017retinal}. Second, the U-Net is more robust than the combination of down-sampling convolution blocks in the discriminator ~\cite{schonfeld2020u}, thus better converges towards the Nash equilibrium \cite{fedus2018many}. The input of discriminator is the concatenation of the fundus image and segmentation map. In order to achieve convergence, the segmenter should produce realistic segmentation maps to minimise the $\mathbb{E}_{x\sim p_{x}(x))}[log(1-D(x,G(x)))]$, corresponding to minimise $L_{G}$ in Figure 3(a). The pixel-level discriminator is trained to maximise the $\mathbb{E}_{y\sim p_{data}(y))}[logD(x,y)]+\mathbb{E}_{x\sim p_{x}(x))}[log(1-D(x,G(x)))]$, i.e., minimising the $L_{d}^{1}$ and $L_{d}^{2}$, where $x$ indicates fundus image and $y$ represents the experts annotation. The loss function over the segmentation network is

\begin{equation}\label{eq2}
Loss_{\textrm{seg}}=\alpha \cdot L_{GAN}+\beta \cdot L_{BCE}+\gamma \cdot L_{MSE}
\end{equation}

\noindent which combines the adversarial loss, binary cross entropy (BCE), and Mean square error (MSE) with weight hyperparameters $\alpha$, $\beta$, and $\gamma$. The BCE and MSE measure the distance between $G(x)$ and $y$, as shown in Figure 3(a).

\subsection{Binary-to-multi-class Fusion Network}

Considering the preferable accuracy in binary segmentation, e.g., artery segmentation shown in Figure \ref{fig1}(e), we devise a binary-to-multi-class fusion network to merge representations of the artery, vein, and multi-class feature to improve the multi-class segmentation performance at the intersections. The whole structure is shown in Figure 3(a). The main segmenter outputs the multi-class vessel map. All three segmenters share the same discriminator. The artery segmentation map $f_{a}$, multi-class feature map $f_{m}$, and vein segmentation map $f_{v}$ are concatenated to generate fused feature maps for the final convolution operation, to generate the multi-class segmentation map.

Additionally, we utilise the deep supervisions in up-sampling stages to avoid gradient vanishing in deep layers ~\cite{lee2015deeply}, thus enhancing the discriminative information. The structure is shown in Figure 3(c). The feature maps in each up-sampling stages $\{{f^{i}, i=1, 2, 3}\}$ are through the structure to respectively obtain $f_{s1}$, $f_{s2}$, $f_{s3}$, to get the deep supervision loss $\left \{ L_{BCE_{i}}, L_{MSE_{i}} \mid i \in (s1, s2, s3) \right \} $. Combining Eq. \ref{eq2}, the fusion loss function combination is

\begin{equation}\label{eq3}
Loss_{\textrm{fusion}}=Loss_{\textrm{seg}} 
+ \sum_{i=1}^3 \frac{1}{2^i} (\beta L_{BCE_{S_i}}+\gamma L_{MSE_{S_i}}),
\end{equation}

\noindent where the $\dfrac{1}{2^i}$ works as the weights for the deep supervision of the three side outputs. The larger weight is allocated to the deeper layer as the risk of gradient vanishing increases.
\section{Experiments}

\subsection{Experiment Setting}

We include three clinical data sets in experiments, the DRIVE-AV~\cite{staal2004ridge,hu2013automated}, LES-AV~\cite{orlando2018towards}, and HRF-AV~\cite{budai2013robust,hemelings2019artery}. DRIVE-AV includes 40 colour retinal fundus images with the size of $(565, 584)$, containing labels for artery, vein, crossing, and uncertainty. The LES-AV data set contains 22 images with the size of $(1620, 1444)$. The HRF-AV data set includes 45 images with the size of $(3504, 2336)$. In DRIVE-AV, 20 images are for training and 20 images for testing. In LES-AV, we sample 11 images for training, and leave 11 images for testing. In HRF-AV, 24 images are used for training and 21 images for testing. 10 percent of training images are used for validation.

\begin{table}[t]
\caption{Multi-class vessel segmentation results on DRIVE-AV, LES-AV, and HRF-AV data sets compared with state-of-the-art methods. Since the TR-GAN is the most competitive compared method, the P-value of Mann–Whitney U test between TR-GAN and the proposed method is calculated to show the statistic significance.}
\label{table1}
\centering
\renewcommand\arraystretch{1.0}
{
\smallskip\begin{tabular}{llllll}
\hline
                  & \multicolumn{5}{c}{\textbf{DRIVE-AV}}\\ \hline
Methods               & Sen        & F1    & ROC     & PR     & MSE   \\ \hline

U-net \cite{ronneberger2015u}            & 53.6$\pm$0.42      & 65.57$\pm$0.58   & 79.84$\pm$0.21       & 65.92$\pm$0.17       & 3.71$\pm$0.01      \\
CS-Net \cite{mou2019cs}      & 52.03$\pm$0.16    & 63.68$\pm$0.20  & 81.96$\pm$0.12    & 63.04$\pm$0.11    & 3.91$\pm$0.02   \\
MNNSA  \cite{ma2019multi}           &59.88$\pm$0.82     & 61.8$\pm$0.33   & 78.89$\pm$0.34       & 65.43$\pm$0.19       & 3.62$\pm$0.02    \\ 
TR-GAN  \cite{chen2020tr}           & 67.84$\pm$1.24    & 65.63$\pm$0.71   & 80.93$\pm$0.57       & 67.44$\pm$0.63       & 3.63$\pm$0.01    \\\hline
Ensemble  &   66.92$\pm$0.67    &  67.13$\pm$0.6   &  81.07$\pm$0.29      &  68.34$\pm$0.17     & 3.49$\pm$0.01 \\ 
Proposed  &  \textbf{69.87$\pm$0.11}      &   \textbf{70.03$\pm$0.03}  &  \textbf{84.13$\pm$0.05}      & \textbf{71.17$\pm$0.03}      & \textbf{3.09$\pm$0.01}   \\ \hline

P-value               & 2.46e-4        & 1.83e-4    & 1.83e-4     & 1.83e-4     & 1.83e-4   \\ \hline

& \multicolumn{5}{c}{\textbf{LES-AV}}\\ \hline

Methods                & Sen    & F1    & ROC     & PR     & MSE    \\ \hline

U-net \cite{ronneberger2015u}                 &  {58.02$\pm$0.49}    & {61.6$\pm$0.41}   & {78.51$\pm$0.23}     &  {65.71$\pm$0.3}  &{2.46$\pm$0.03} \\
CS-Net \cite{mou2019cs}       & {49.88$\pm$0.88}    & {58.58$\pm$0.83}   & {74.12$\pm$0.16}     &  {64.99$\pm$0.14}  &{2.51$\pm$0.04} \\
MNNSA   \cite{ma2019multi}             & {51.31$\pm$0.58}     & {57.16$\pm$0.22}   &{76.75$\pm$0.33}     & {58.86$\pm$0.22}  &{2.82$\pm$0.05}\\ 
TR-GAN   \cite{chen2020tr}             & {59.33$\pm$0.72}     & {61.14$\pm$0.57}   &{77.57$\pm$0.19}     & {65.4$\pm$0.26}  &{2.54$\pm$0.07}\\\hline
Ensemble  & 58.81$\pm$0.58    &   62.51$\pm$0.34  &   78.58$\pm$0.28     &    66.31$\pm$0.13   & 2.59$\pm$0.07 \\
Proposed      &  \textbf{62.94$\pm$0.93}      &  \textbf{66.69$\pm$0.47}  &   \textbf{81.03$\pm$0.04}   &   \textbf{68.71$\pm$0.47} &  \textbf{2.17$\pm$0.05}\\\hline

P-value               & 1.83e-4        & 1.83e-4    & 1.83e-4     & 1.83e-4     & 1.83e-4    \\ \hline

& \multicolumn{5}{c}{\textbf{HRF-AV}}\\ \hline

Methods                & Sen       & F1    & ROC     & PR     & MSE    \\ \hline

U-net \cite{ronneberger2015u}                 &  {63.43$\pm$1.12}        & {67.12$\pm$0.85}     &  {79.98$\pm$0.77}  &{72.2$\pm$0.42} & {2.49$\pm$0.01}\\
CS-Net \cite{mou2019cs}       & {62.5$\pm$1.47}     & {66.07$\pm$0.66}   & {79.12$\pm$0.86}     &  {69.18$\pm$0.86}  &{2.64$\pm$0.01} \\
MNNSA   \cite{ma2019multi}             & {62.81$\pm$0.92}    & {64.3$\pm$0.63}   &{78.83$\pm$0.79}     & {69.06$\pm$0.46}  &{2.78$\pm$0.02}\\ 
TR-GAN   \cite{chen2020tr}             & {63.14$\pm$0.97}     & {67.5$\pm$0.39}   &{80.67$\pm$0.74}     & {72.45$\pm$0.38}  &{2.37$\pm$0.02}\\\hline
Ensemble  & 64.29$\pm$1.39    &   68.58$\pm$0.87  &   80.33$\pm$0.41     &    71.09$\pm$0.59   & 2.24$\pm$0.01 \\
Proposed      &  \textbf{67.68$\pm$1.57}     &  \textbf{71.7$\pm$0.44}  &   \textbf{83.44$\pm$0.75}   &   \textbf{73.96$\pm$0.31} &  \textbf{1.9$\pm$0.01}\\ \hline

P-value              & 1.83e-4        & 1.83e-4    & 1.83e-4     & 1.83e-4     & 1.83e-4   \\ \hline

\end{tabular}}
\end{table}

\begin{table}[t]

	\centering
		\caption{The ablation study of binary-to-multi-class fusion network. Adversarial segmentation network (\textit{Seg}), deep supervision (\textit{Deep}), and binary-to-multi-class fusion (\textit{BF}). The metrics AUC-ROC (\textit{ROC}) and F1-score (\textit{F1}). Growth is shown in bracket.}
		 \label{table2}
		{
			\smallskip\begin{tabular}{cccclllllllll}
				\hline
				\multicolumn{3}{c}{Components}                                                           & \multicolumn{2}{c}{DRIVE-AV}          & \multicolumn{2}{c}{LEV-AV}    & \multicolumn{2}{c}{HRF-AV}     \\ \hline
				 \textit{Seg} & \textit{Deep} & \textit{BF} & \textit{ROC} & \textit{F1} & \textit{ROC} & \textit{F1} & \textit{ROC} & \textit{F1}\\ \hline
					                 &                   &                      &        79.84     &       65.57        &     78.51         &       61.6       &     79.98     &  67.12 \\
					       $\surd$           &                   &                      &   $81.77_{(1.93)}$          &        $67.45_{(1.88)}$       &        $79.44_{(0.93)}$       &       $63.78_{(2.18)}$        &   $81.68_{(1.7)}$   &  $69.41_{(2.29)}$ \\
					        $\surd$           &    $\surd$                &                      & $82.3_{(0.53)}$           &      $68.52_{(1.07)}$          &       $79.95_{(0.51)}$        &     $64.93_{(1.15)}$         &       $82.31_{(0.63)}$   &  $70.16_{(0.75)}$    \\
					        $\surd$           &    $\surd$                &      $\surd$                 &    $84.13_{(1.83)}$          & $70.03_{(1.51)}$       &      $81.03_{(1.08)}$         & $66.69_{(1.76)}$    &        $83.44_{(1.13)}$   &  $71.7_{(1.54)}$ \\\hline

		\end{tabular}}

\end{table}

For implementation and training details, the images in DRIVE-AV are zero-padded to the size of $(592, 592)$. The images in LES-AV and HRF-AV are respectively resized to $(800, 720)$ and $(880, 592)$ in the training, for computational consideration~\cite{galdran2019uncertainty,li2020joint}. In validation, the images are resized back to the original size for metrics calculation. We set learning rate 0.0008, and batch size 2 across experiments. $\alpha$, $\beta$, and $\gamma$ in Eq. \ref{eq2} are set at 0.08, 1.1, and 0.5 respectively. The hyper-parameter search is in Supplementary Materials section 1. Total training time of a model is approximately 12 hours for 1500 epochs on one Tesla T4.

For evaluation metrics, the retinal vessel multi-task segmentation is an unbalanced task. Some metrics like accuracy and specificity are always high, thus losing practical value. We employ AUC (Area Under The Curve)-ROC (Receiver Operating Characteristics), sensitivity, F1-score, AUC-PR (Precision-Recall), and MSE to comprehensively evaluate the methods. For example, the F1-score computation is $F=(n_{a} \cdot F_{a}+ n_{v} \cdot F_{v} + n_{u} \cdot F_{u})/(n_{a}+n_{v}+n_{u})$, where $n_{a}$, $n_{v}$, and $n_{u}$ respectively represent the pixels number belonging to artery, vein, and uncertainty in label. $F_{a}$, $F_{v}$, and $F_{u}$ are the F1-score scores in each binary measurement, e.g., artery pixels versus all other pixels. The Mann–Whitney U test is employed to test the statistic significance of the difference across methods. For fair comparison, we implemented all baselines with the same pre-processing, input size, and evaluation methods.

\subsection{Experiment Results}

 \subsubsection{Comparing with most recent methods} 
\begin{figure}[t]
\centering
\includegraphics[width=0.8\columnwidth]{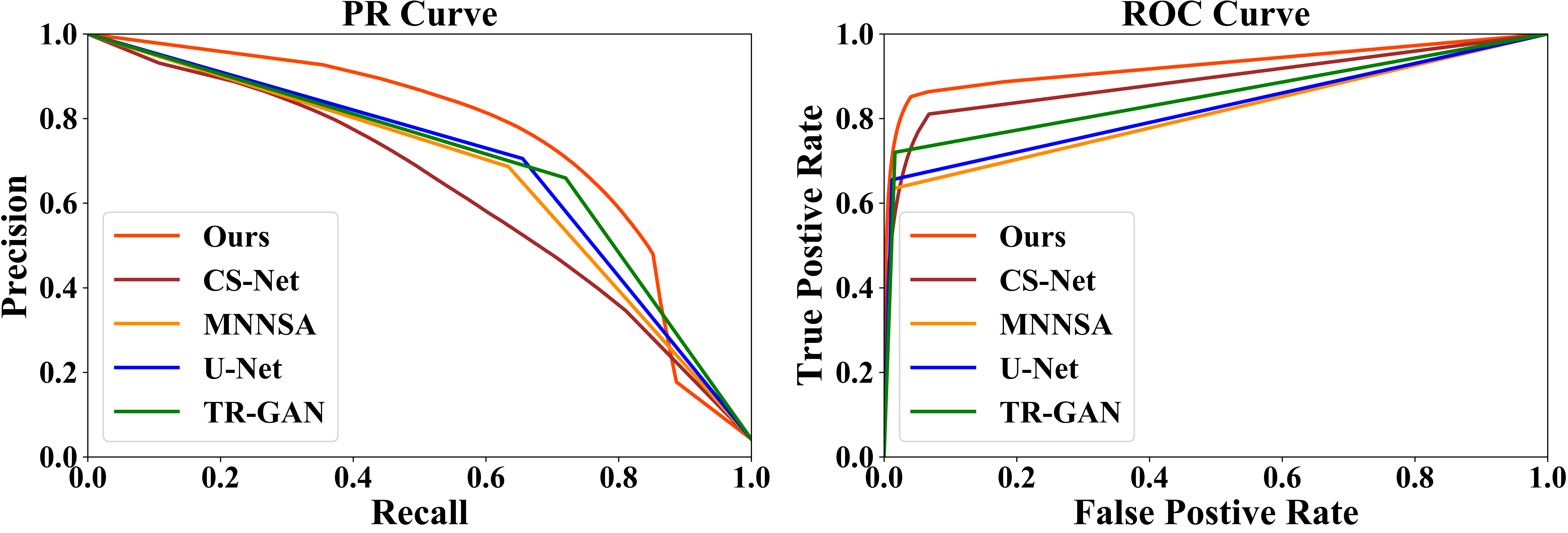}
\caption{ROC and PR curves of various multi-class segmentation methods.}\label{fig4}
\end{figure}

The compared methods CS-Net~\cite{mou2019cs}, MNNSA~\cite{ma2019multi}, and TR-GAN~\cite{chen2020tr} are recent related works. The U-Net is a baseline. We list the overall multi-class vessel segmentation performance in Table \ref{table1}. The P-value of Mann–Whitney U test is smaller than 0.05, showing statistic significance. The ROC and PR curves on the DRIVE-AV test are shown in Figure \ref{fig4}. Based on the results, the proposed method enhances performance, e.g., F1-scores respectively increase by 4.4\%, 5.09\%, 4.2\% in DRIVE-AV, LES-AV, and HRF-AV. For all segmentation maps, the reader is referred to Supplementary Materials section 2.

\begin{figure}[t]
\centering
\includegraphics[width=\columnwidth]{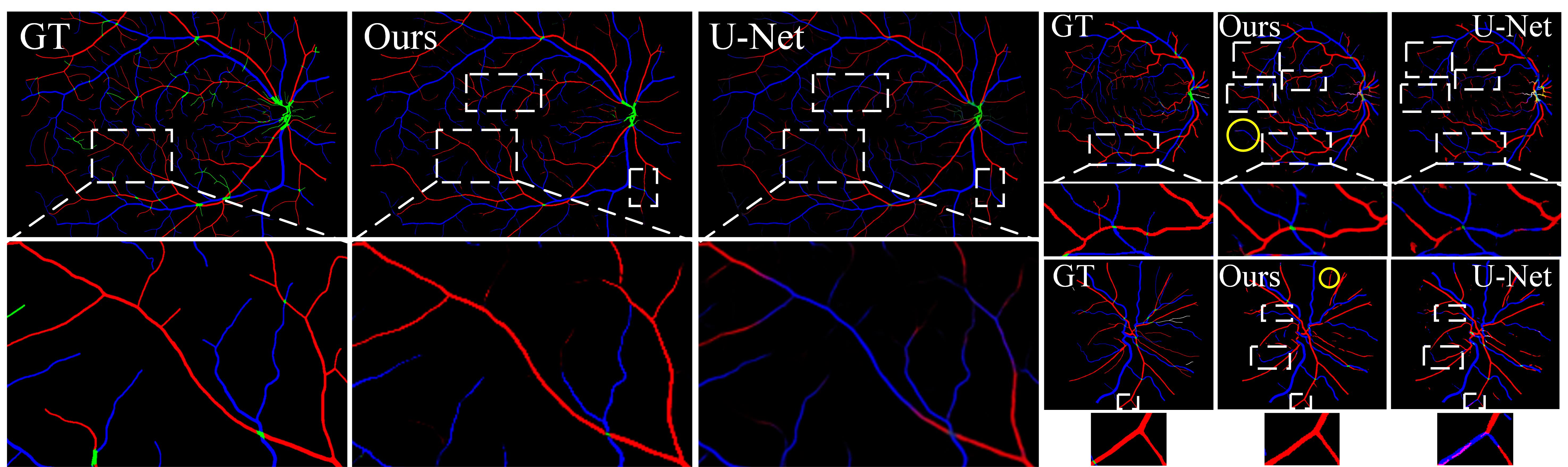}
\caption{The visualisation results with the proposed method and U-Net. The White dotted boxes show solved intra-segment misclassification, while yellow circle shows some residual issues. Better view with zoom in.} \label{fig5}
\end{figure}

\subsubsection{Ablation Study}
We evaluate the contribution of each component from the binary-to-multi-class fusion network, including with and without the adversarial segmentation network, deep supervision, and binary-to-multi-class fusion. As shown in Table \ref{table2}, the adversarial segmentation network performs better due to skip-connection based merging and the pixel-level adversarial learning, when compared with vanilla U-Net \cite{ronneberger2015u} in first line. The binary-to-multi-class fusion network improves the overall metrics, with the largest improvement 1.83\% in AUC-ROC and 1.76\% in F1-score. Observing the benefits brought by BF, the benefit of introducing artery and vein binary branches is highlighted. With the ``Ensemble" method in Table \ref{table1}, we separately train two binary adversarial segmentation networks with artery and vein labels, and then concatenate the artery and vein maps to get the multi-class segmentation map, aiming to verify the effectiveness of our learning based merging strategy.

\section{Discussion}

Based on the results from the ablation experiment, the adversarial segmentation network and the binary-to-multi-class fusion network contribute most of the performance improvement as they are specifically dedicated to the intra-segment misclassification. The deep supervision also improves the performance albeit no more than 1.15\% in F1-score. Although both qualitative and quantitative results suggest that the intra-segment misclassification has been noticeably reduced, there were still cases where the problem persisted albeit to a much reduced extent, as shown in the yellow circles in Figure \ref{fig5}. In particular, branches with small vessels are generally more prone to the intra-segment misclassification.

We would explore two aspects for future direction. Firstly, topological regularisation may be embedded in the model via various approaches, such as post-processing, customised convolution filter, and the feature merging between topological and representative structures. Secondly, it may be interesting to investigate the vessel segmentation problem in low-contrast background. Moreover, the proposed methods may be further evaluated by embedding the downstream clinical applications, such as the hypertension, diabetes, and atherosclerosis early diagnosis and grading.

In this work, we first identify a key point for improving the multi-class segmentation task in a real clinical application, that is the intra-segment misclassification caused by inter-class vessel intersections. Based on the hypothesis that the binary branch segmenter offers representation to the multi-class main segmenter to enhance the performance around the inter-class intersections, a binary-to-multi-class fusion network is proposed. According to experiments on three clinical data sets, our hypothesis is verified as the proposed network achieves new state-of-the-art performance.

\subsubsection{Acknowledgements}
This work is supported by EPSRC grants EP/M020533/1 EP/R014019/1 and EP/V034537/1 as well as the NIHR UCLH Biomedical Research Centre. Dr Keane is supported by a Moorfields Eye Charity Career Development Award (R190028A) and a UK Research $\&$ Innovation Future Leaders Fellowship (MR/T019050/1). Dr Jacob is funded by a Wellcome Trust Clinical Research Career Development Fellowship 209553/Z/17/Z.

%
%
\bibliographystyle{splncs04}
\bibliography{reference}

\title{Appendix: Learning to Address Intra-segment Misclassification in Retinal Imaging}
%
%
\author{}
\authorrunning{}
%
\institute{
}
\maketitle

\section{Hyper-parameters Optimisation}

\begin{figure*}[]
\centering
\includegraphics[width=1\columnwidth]{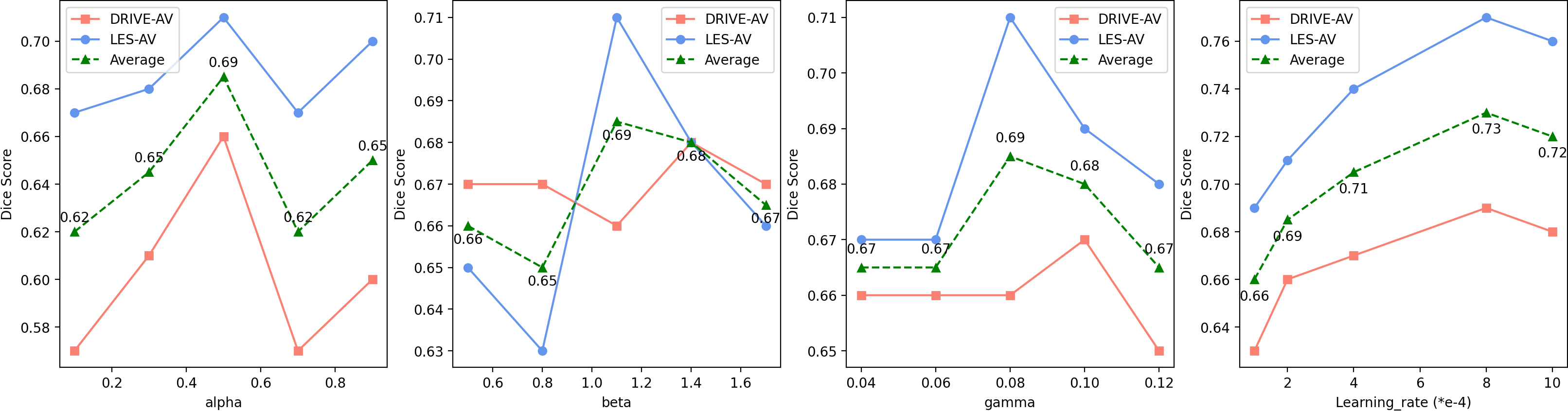}
\caption{Hyper-parameters searching on the three loss weights $\alpha$, $\beta$, $\gamma$, in Eq. (2), and learning rate, respectively.  The red line represents performance on DRIVE-AV and the blue one represents LES-AV. The green dotted line is the mean value between red and blue. When evaluation on one of the hyper-parameters, the other three are with initial searching value ($\alpha$ is 0.5, $\beta$ is 1.1, $\gamma$ is 0.08, and learning rate is 0.0002).}\label{appendixfig1}
\end{figure*}

\section{Visualisation Results}

\begin{figure}[h]
\centering
\includegraphics[width=1\columnwidth]{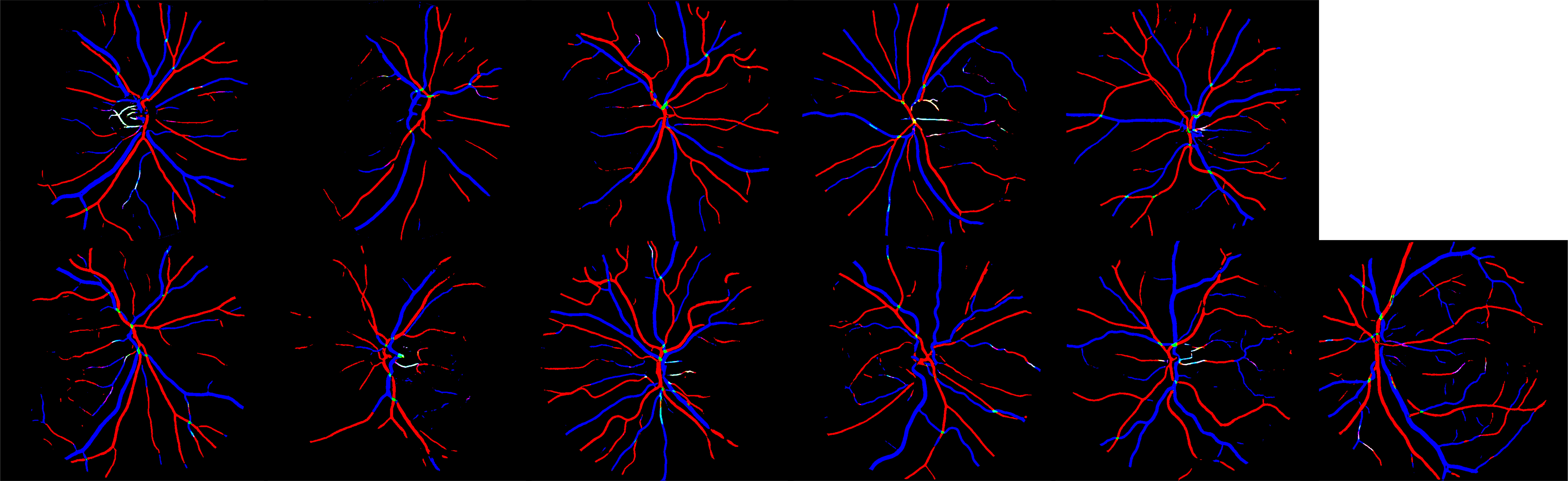}
\caption{Segmentation of the binary-to-multi fusion network on LES-AV}\label{appendixfig2}
\end{figure}

\begin{figure}[h]
\centering
\includegraphics[width=1\columnwidth]{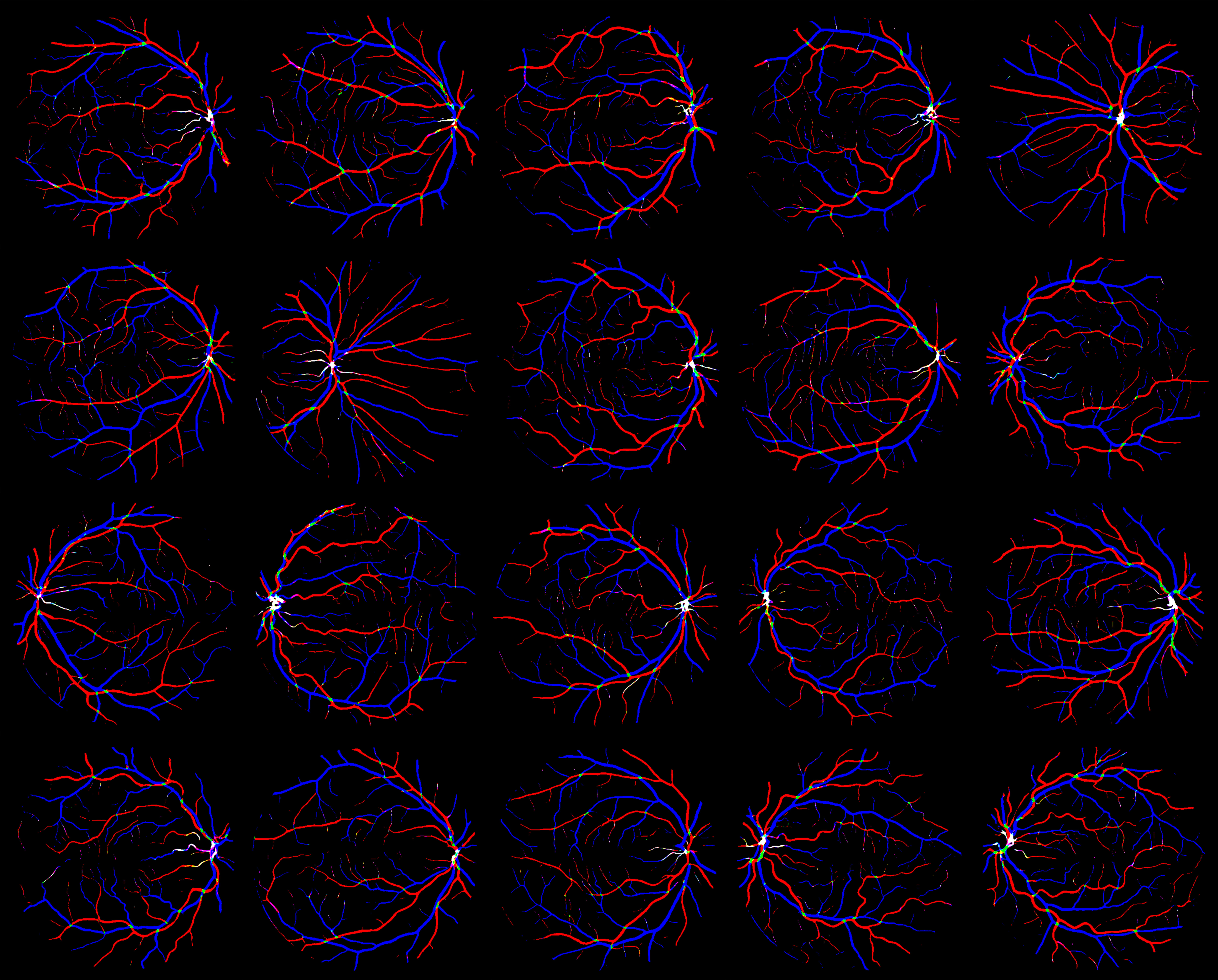}
\caption{Segmentation of the binary-to-multi fusion network on DRIVE-AV}\label{appendixfig3}
\end{figure}

\begin{figure}[h]
\centering
\includegraphics[width=1\columnwidth]{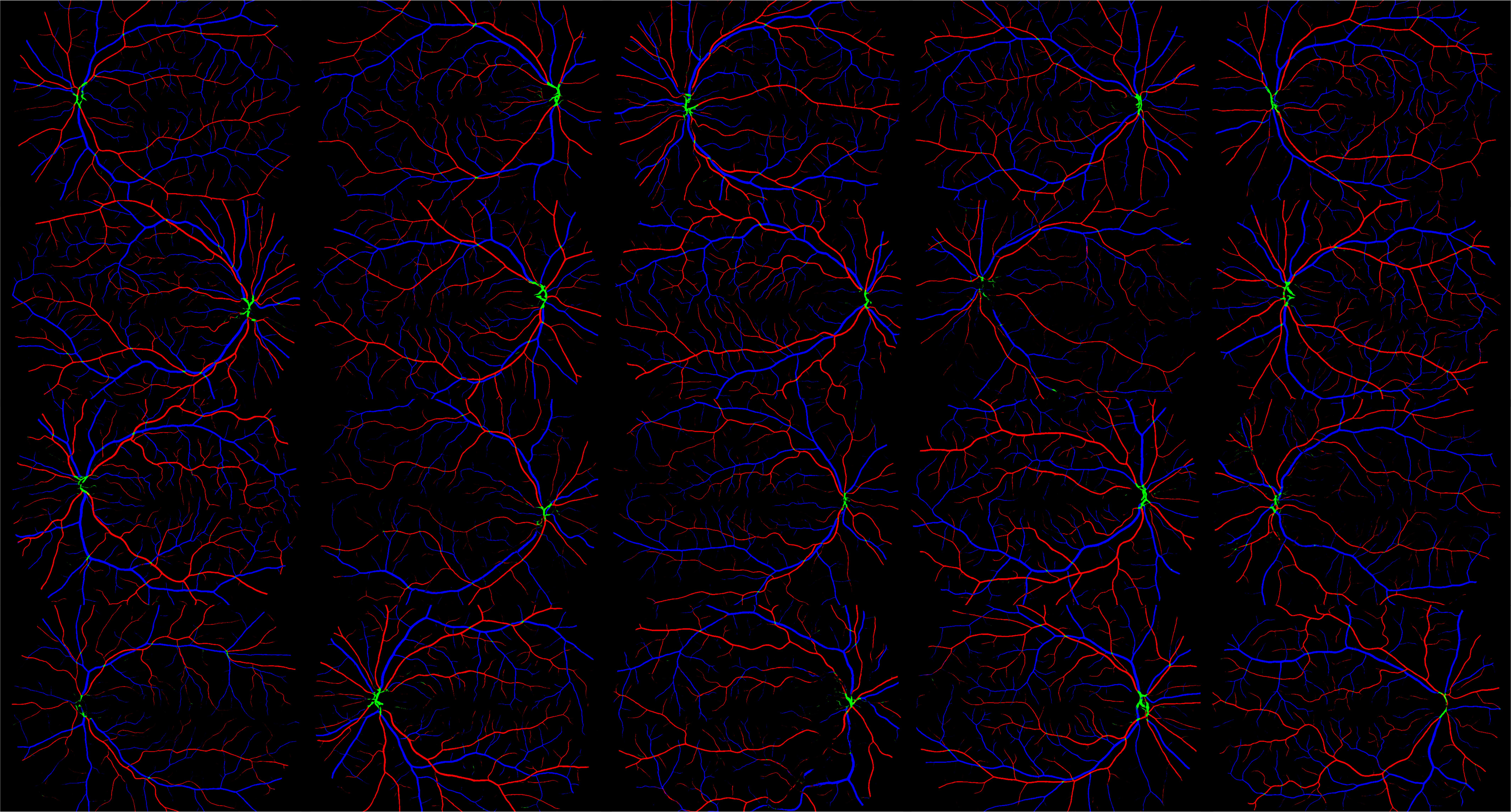}
\caption{Segmentation of the binary-to-multi fusion network on HRF-AV}\label{appendixfig4}
\end{figure}

%
%

\end{document}